\documentclass{sig-alternate}

\usepackage{color}
\usepackage{algorithm, algpseudocode}
\algrenewcommand\algorithmicrequire{\textbf{\quad Input:}}
\algrenewcommand\algorithmicensure{\textbf{\quad Output:}}

\newtheorem{theorem}{Theorem}[section]
\newtheorem{lemma}[theorem]{Lemma}
\newtheorem{claim}[theorem]{Claim}
\newtheorem{corollary}[theorem]{Corollary}
\newtheorem{definition}[theorem]{Definition}
\newcommand{\qedsymb}{\hfill{\rule{2mm}{2mm}}}
\newenvironment{exrefproof}[2]{\begin{trivlist} \item[\hspace{\labelsep}{\bf \noindent Proof of #1 #2.\/}] }{\qedsymb\end{trivlist}} %
\newcommand{\bs}[1]{\boldsymbol{#1}}
\newcommand{\bbR}{\mathbb{R}}
\newcommand{\bbN}{\mathbb{N}}
\newcommand{\sfrac}[2]{{^{#1}} \hspace{-0.1cm} / \hspace{-0.05cm} {_{#2}}}

\begin{document}

\conferenceinfo{SPAA'07,} {June 9--11, 2007, San Diego, California, USA.}
\CopyrightYear{2007}
\crdata{978-1-59593-667-7/07/0006}

% TITLE %%%%%%%%%%%%%%%%%%%%%%%%%%%%%%%%%%%%%%%%%%%%%%%%%%%%%%%%%%%%
%\begin{titlepage}
\title{Truthful Unsplittable Flow for Large Capacity Networks}
%\author{Yossi Azar\thanks{ {\tt azar@post.tau.ac.il}. School of
%        Computer Science, Tel-Aviv University, Tel-Aviv, 69978,
%        Israel. Supported in part by the German-Israeli Foundation
%        and by the Israel Science Foundation.}
%        \and
%        Iftah Gamzu\thanks{ {\tt iftgam@post.tau.ac.il}. School of
%        Computer Science, Tel-Aviv University, Tel-Aviv, 69978,
%        Israel. }
%        \and
%        Shai Gutner\thanks{ {\tt gutner@post.tau.ac.il}. School of
%        Computer Science, Tel-Aviv University, Tel-Aviv, 69978,
%        Israel.}
%    }
%\date{}

\numberofauthors{3}
\author{
\alignauthor Yossi Azar \titlenote{Research supported in part by
the Israel Science Foundation and by the German-Israeli Foundation.}\\
\affaddr{Microsoft Research, Redmond and}\\
\affaddr{Tel-Aviv University}\\
\affaddr{Tel-Aviv, 69978, Israel}\\
\email{azar@tau.ac.il}
\alignauthor{Iftah Gamzu}\\
\affaddr{School of Computer Science}\\
\affaddr{Tel-Aviv University}\\
\affaddr{Tel-Aviv, 69978, Israel}\\
\email{iftgam@post.tau.ac.il}
\alignauthor Shai Gutner \titlenote{This paper forms part of a
Ph.D. thesis written by the author under the supervision of Prof.
N. Alon and Prof. Y. Azar in Tel Aviv University.}\\
\affaddr{School of Computer Science}\\
\affaddr{Tel-Aviv University}\\
\affaddr{Tel-Aviv, 69978, Israel}\\
\email{gutner@tau.ac.il}
}

\maketitle
%\thispagestyle{empty}

% ABSTRACT %%%%%%%%%%%%%%%%%%%%%%%%%%%%%%%%%%%%%%%%%%%%%%%%%%%%%%%%%
\begin{abstract}
The \emph{unsplittable flow problem} is one of the most
extensively studied optimization problems in the field of
networking. An instance of it consists of an edge capacitated
graph and a set of connection requests, each of which is
associated with source and target vertices, a demand, and a value.
The objective is to route a maximum value subset of requests
subject to the edge capacities. It is a well known fact that as
the capacities of the edges are larger with respect to the maximal
demand among the requests, the problem can be approximated better.
In particular, it is known that for sufficiently large capacities,
the integrality gap of the corresponding integer linear program
becomes $1+\epsilon$, which can be matched by an algorithm that
utilizes the randomized rounding technique.

In this paper, we focus our attention on the large capacities
unsplittable flow problem in a game theoretic setting. In this
setting, there are selfish agents, which control some of the
requests characteristics, and may be dishonest about them. It is
worth noting that in game theoretic settings many standard
techniques, such as randomized rounding, violate certain
monotonicity properties, which are imperative for truthfulness,
and therefore cannot be employed. In light of this state of
affairs, we design a monotone deterministic algorithm, which is
based on a primal-dual machinery, which attains an approximation
ratio of $\frac{e}{e-1}$, up to a disparity of $\epsilon$ away.
This implies an improvement on the current best truthful
mechanism, as well as an improvement on the current best
combinatorial algorithm for the problem under consideration.
Surprisingly, we demonstrate that any algorithm in the family of
reasonable iterative path minimizing algorithms, cannot yield a
better approximation ratio. Consequently, it follows that in order
to achieve a monotone PTAS, if exists, one would have to exert
different techniques. We also consider the large capacities
\textit{single-minded multi-unit combinatorial auction problem}.
This problem is closely related to the unsplittable flow problem
since one can formulate it as a special case of the integer linear
program of the unsplittable flow problem. Accordingly, we obtain a
comparable performance guarantee by refining the algorithm
suggested for the unsplittable flow problem.
\end{abstract}

%\bigskip {\bf Remark}: submitted for regular presentation.

%\end{titlepage}

%\newpage


% INTRO %%%%%%%%%%%%%%%%%%%%%%%%%%%%%%%%%%%%%%%%%%%%%%%%%%%%%%%%%%%%
\section{Introduction}

\noindent {\bf The problems.} We study the \emph{unsplittable flow
problem}. As input to this problem, we are given a directed or
undirected graph $G = (V, E)$, such that $n = |V|$, $m = |E|$, and
every edge $e \in E$ has a positive capacity $c_{e}$. An
additional ingredient of the input is a set $\mathcal{R}$ of
connection requests, in which every request $r \in \mathcal{R}$ is
characterized by a quadruple $(s_{r}, t_{r}, d_{r}, v_{r})$ such
that $s_{r}$ and $t_{r}$ are the respective \textit{source} and
\textit{target} vertices of the request, $d_{r}$ is the positive
\textit{demand} associated with the request, and $v_{r}$ is the
positive \textit{value} or profit gained as a result of allocating
the request. The objective is to select a maximum value subset of
requests $S \subseteq \mathcal{R}$, along with a path for each
selected request, so that all the requests in $S$ can
simultaneously route their demand along the corresponding path,
while preserving the capacity constraints. Denoting by $B =
\min_{e} \{ c_{e} \} / \max_{r} \{ d_{r} \}$ the ratio between the
minimal capacity of an edge and the maximal demand among the
requests, the problem is referred to as the \textit{B-bounded
unsplittable flow problem}. Since one can normalize both the
demands of the requests and the capacities of the edges, the
B-bounded unsplittable flow problem can be equivalently defined to
have $d_{r} \in (0,1]$ for every request $r$, and $B = \min_{e} \{
c_{e} \}$. Note that we shall use the latter definition throughout
this paper.

We also consider the \textit{single-minded multi-unit
combinatorial auction problem}. This problem is closely related to
the unsplittable flow problem since one can formulate it as a
special case of the integer linear program of the unsplittable
flow problem. An instance of it consists of a set $U$ of $m$
non-identical items, such that item $u \in U$ has a positive
integer multiplicity $c_{u}$. The input also consists of a set
$\mathcal{R}$ of requests, in which every request $r \in
\mathcal{R}$ is characterized by a pair $(U_{r}, v_{r})$ such that
$U_{r} \subseteq U$ is an items bundle, which is the \emph{demand}
associated with the request, and $v_{r}$ is the positive
\textit{value} gained as a result of allocating the bundle. The
goal is to select a maximum value subset $S \subseteq
\mathcal{R}$, so that every item $u \in U$ appears in at most
$c_{u}$ bundles of requests in $S$. Denoting by $B = \min_{u} \{
c_{u} \}$ the minimum multiplicity of an item, the problem is
referred to as the \textit{B-bounded multi-unit combinatorial
auction problem}.

\smallskip \noindent {\bf The setting.} In the present paper, we
study the \emph{$\Omega(\ln m)$-bounded unsplittable flow
problem}, and the \emph{$\Omega(\ln m)$-bounded multi-unit
combinatorial auction problem} from a \emph{mechanism design}
\cite{NisanR99} point of view. In this game-theoretic setting,
some characteristics of the requests, which are henceforth
referred to as the \emph{type} of the requests, are controlled by
\emph{selfish agents}. An agent is selfish in a sense that it
might declare a fallacious type in order to manipulate the
algorithm in a way that will maximize its own utility. Our goal is
to design mechanisms, which are referred to as \emph{incentive
compatible} or \emph{truthful}, in which each agent's best
strategy is always to reveal the true type of the request that it
controls, regardless of the other requests types, and regardless
of the way that the other agents decide to declare their requests
types. In particular, we aim to devise \emph{monotone} algorithms
which are, roughly speaking, equivalent to truthful mechanisms.
Note that in the unsplittable flow problem, the type of a request
is its demand and value, whereas in the multi-unit combinatorial
auction problem the type of a request is its value. Also note that
the other characteristics of the request, e.g.\ the source and
target vertices in the unsplittable flow problem, are assumed to
be known and thus, the agent cannot be untruthful about them.

\smallskip \noindent {\bf The motivation.} One of the closely related
problems to the unsplittable flow problem is the
\emph{multicommodity flow problem}. Since the multicommodity flow
problem can be modeled by the relaxation of the integer linear
program of the unsplittable flow problem, it may be considered as
its fractional version. It is well known that the integrality gap
of the integer linear program of the unsplittable flow problem
becomes $1 + \epsilon$ when the ratio between the minimal capacity
of an edge and the maximal demand among the requests is
sufficiently large. Consequently, it is conjectured that in such
case, the performance of algorithms for the fractional and
integral versions would be similar. Specifically, since the
multicommodity flow problem admits a monotone PTAS by
combinatorial primal-dual based algorithms
\cite{GargK98,Fleischer99}, one may expect that an integral
version of these monotone PTAS would yield a monotone PTAS for the
unsplittable flow problem. In the following, we refute this
perception. In particular, we design a primal-dual based monotone
algorithm for the $\Omega(\ln m)$-bounded unsplittable flow
problem that attains the best possible approximation ratio with
respect to any \emph{reasonable iterative path minimizing}
algorithm\footnote{The family of \emph{reasonable iterative path
minimizing} algorithms is formally defined in
Subsection~\ref{cha:UfpInapprox}.}, and is not a PTAS.
Nevertheless, This algorithm still improves upon the best previous
known result \cite{BriestKV05}.

\subsection{Our results}

\noindent {\bf Monotone deterministic algorithms.} We describe a
monotone deterministic algorithm based on a primal-dual approach
for the $\Omega(\ln m)$-bounded unsplittable flow problem, which
obtains an approximation ratio that approaches $\frac{e}{e-1}
\approx 1.58$. This result implies a corresponding incentive
compatible mechanism. In addition, we show that the aforesaid
algorithm can be attuned for the $\Omega(\ln m)$-bounded
multi-unit combinatorial auction problem, and attain a comparable
approximation ratio, i.e.\ $\frac{e}{e-1}$-approximation. These
results improve over the approximation guarantee suggested by
Briest et al.\ \cite{BriestKV05} for both problems, which
approaches $e \approx 2.71$.

\smallskip \noindent {\bf Deterministic lower bounds.}
We prove that any algorithm for the $\Omega(\ln m)$-bounded
unsplittable flow problem, which is part of the reasonable
iterative path minimizing algorithms family, cannot yield an
approximation guarantee that is better than $\frac{e}{e-1} -
o(1)$. This implies, on the one hand, that the analysis of our
algorithm is tight, and on the other hand, that to achieve a
monotone deterministic PTAS, if exists, one would have to employ
different techniques. Additionally, we reinforce this
inapproximability result by demonstrating that even if we ease the
problem setting, e.g.\ assume that the minimal capacity of an edge
is arbitrarily large, still no reasonable iterative path
minimizing algorithm can attain PTAS. Correspondingly, we also
establish a lower bound of $\sfrac{4}{3}$ on the approximation
ratio of any reasonable iterative bundle minimizing algorithm for
the $\Omega(\ln m)$-bounded multi-unit combinatorial auction
problem.

\smallskip \noindent {\bf A deterministic $\bs{(1 + \epsilon)}$-approximation algorithm.}

\noindent
We study the $\Omega(\ln m)$\emph{-bounded unsplittable flow with
repetitions problem}, which is a variant of the $\Omega(\ln
m)$-bounded unsplittable flow problem in which one is allowed to
satisfy every request multiple times using possibly multiple
paths. In contrast with our prior findings, we demonstrate that
this version admits a deterministic primal-dual based algorithm,
which yields an $(1 + \epsilon)$-approximation.

\subsection{Related work}

The unsplittable flow problem, and its fractional variant, the
multicommodity flow problem, has been given an extensive attention
in recent years, both from an algorithmic point of view and from a
game-theoretic one. These fundamental optimization problems model
a diverse collection of applications in network routing, parallel
computing, and even in VLSI layout. Obviously, the fractional
problem is easier. In particular, it is known to admit an optimal
solution by linear programming, and an $(1 +
\epsilon)$-approximate solution by combinatorial algorithms, see
e.g.\ \cite{GargK98,Fleischer99}. In contrast with the fractional
problem, approximating the integral problem is hard. Guruswami and
Talwar \cite{ECCC-TR06-141} have recently showed that the directed
version of the problem is $n^{\Omega(1/B)}$-hard to approximate
unless $\mathrm{NP} \subseteq \mathrm{BPTIME}(n^{O(\log\log n)})$,
where $B = O(\log n / \log\log n)$ is the minimal capacity of an
edge. This result extended the prominent result of Guruswami et
al.\ \cite{GuruswamiKRSY99}, which states that if $B = 1$, it is
$\mathrm{NP}$-hard to approximate the problem to within a factor
of $O(n^{\sfrac{1}{2} - \epsilon})$. Respectively, when the graph
is undirected, Andrews et al.\ \cite{AndrewsCZ05} established an
$(\log n)^{\Omega(1/B)}$-hardness for any $B = O(\log\log n /
\log\log\log n)$, under the assumption that $\mathrm{NP}
\nsubseteq \mathrm{ZPTIME}(n^{\mathrm{polylog}(n)})$.
Nevertheless, when $B$ is sufficiently large, e.g.\ $B =
\Omega(\ln m)$, the integrality gap of the integer linear program
of the problem becomes $1 + \epsilon$, which can be matched by an
algorithm that utilizes the randomized rounding technique
\cite{RaghavanT87,Raghavan88,Srinivasan99}. Unfortunately, this
standard technique violates certain monotonicity properties, which
are imperative for truthfulness and thus, cannot be directly used
in the presence of selfish agents to obtain a truthful mechanism.
Accordingly, until recently, the known truthful results for the
$\Omega(\ln m)$-bounded unsplittable flow problem only guaranteed
approximation ratios that were at least logarithmic in the size of
the graph \cite{AzarR06,BartalGN03,AwerbuchAM03}. Briest et al.\
\cite{BriestKV05} seem to have been the first to propose a
constant factor approximation algorithm. Essentially, they
designed a monotone primal-dual based algorithm, which was
motivated by the novel work of Garg and K{\"o}nemann
\cite{GargK98} on the fractional problem, that achieves an
approximation guarantee that approaches $e$.

The research of the single-minded multi-unit combinatorial auction
problem, which is closely related to the unsplittable flow
problem, yielded similar results. Bartal et al.\ \cite{BartalGN03}
showed that approximating the problem to within a factor of
$O(m^{1 / (B+1)})$ is $\mathrm{NP}$-hard, where $B$ is the minimum
multiplicity of an item. Yet, when $B = \Omega (\ln m)$, the
integrality gap of the corresponding integer linear program
becomes $1 + \epsilon$. Accordingly, Archer et al.\
\cite{ArcherPTT03}, and Lavi and Swamy \cite{LaviS05} devised
truthful $(1 + \epsilon)$-approximation mechanisms. However, these
mechanisms are truthful only in a probabilistic sense and hence,
the best known deterministic truthful result for the $\Omega(\ln
m)$-bounded multi-unit combinatorial auction problem is by Briest
et al.\ \cite{BriestKV05}, which attains $e$-approximation.

% PRELIMINARIES %%%%%%%%%%%%%%%%%%%%%%%%%%%%%%%%%%%%%%%%%%%%%%%%%%%%
\section{Preliminaries}

In what follows, we present the notions of \emph{monotonicity} and
\emph{exactness}, and then turn to describe a characterization
that reduces the goal of designing truthful mechanisms to that of
designing monotone and exact algorithms. Remark that the
illustrated terms are presented in the context of the problems
under considerations and hence, the keen reader may refer to
Lehmann et al.\ \cite{LehmannOS02} or Briest et al.\
\cite{BriestKV05} for more formal and comprehensive definitions of
the underlying concepts.

\begin{definition}
An algorithm $\mathcal{A}$ for the unsplittable flow problem is
said to be \emph{monotone} w.r.t.\ the demand and value of a
request $r \in \mathcal{R}$, if it satisfies the following
property: if algorithm $\mathcal{A}$ selects $r$ when its demand
is ${d}_{r}$ and its value is $v_{r}$ then algorithm $\mathcal{A}$
would have selected $r$ if its demand was $\tilde{d_{r}} \leq
d_{r}$, its value was $\tilde{v_{r}} \geq v_{r}$, and the demands
and values of all the other requests were fixed.
\end{definition}

\begin{definition}
An algorithm $\mathcal{A}$ for the unsplittable flow problem is
called \emph{exact}, if it allocates the exact demand of every
request selected, and does not allocate anything otherwise.
\end{definition}

\begin{theorem} \label{th:truthful} \emph{(\cite{LehmannOS02,BriestKV05})}
If algorithm $\mathcal{A}$ for the unsplittable flow problem is
monotone and exact w.r.t.\ the demand and value of every request
then there exists a corresponding truthful mechanism. In addition,
this mechanism can be efficiently computed using algorithm
$\mathcal{A}$.
\end{theorem}

Note that similar definitions can analogously be made for the
single-minded multi-unit combinatorial auction problem. The only
exception is that the monotonicity property, and the
characterization theorem are only defined with respect to the
value of every request, i.e.\ the demand terms need to be cast
off.

% UNSPLITTABLE FLOW PROBLEM %%%%%%%%%%%%%%%%%%%%%%%%%%%%%%%%%%%%%%%%
\section{Unsplittable Flow Problem} \label{cha:UfpSect}

\subsection{The algorithm} \label{cha:UfpAlgoSect}
In this subsection, we devise a deterministic monotone algorithm
for the $\Omega(\ln m)$-bounded unsplittable flow problem, which
achieves an approximation ratio that approaches $\frac{e}{e-1}$.
Our algorithm is based on a primal-dual machinery. Accordingly, we
present in Figure \ref{cap:UFPILPandDualLP}, the primal-dual
formulation of the unsplittable flow problem. This will be later
used to motivate the algorithm.

\begin{figure}[!hbt]
\centering
\begin{tabular}{|p{7.0cm}|}
\hline
\begin{eqnarray*}
\nonumber \\[-6ex]
\textbf{\textrm{max}} \;\quad \sum_{r \in \mathcal{R}} v_{r} \cdot \Big(\sum_{s \in S_{r}} x_{s}\Big)\label{eq:ILPgoal}\\\\[-2ex]
    \nonumber\\
\textbf{\textrm{s.t.}} \quad \sum_{s \in S | e \in s} x_{s}   d_{s} \leq c_{e}\quad&  \forall e \in E\label{eq:ILPedges}\\
    \sum_{s \in S_{r}} x_{s} \leq 1\quad\quad\quad& \forall r \in \mathcal{R} \label{eq:ILPflows}\\
    x_{s} \in \{ 0,1 \}\quad\quad\quad& \forall s \in S \label{eq:ILPvars} \\[-6ex]
\nonumber
\end{eqnarray*} \\
\hline
\begin{eqnarray*}
\nonumber \\[-6ex]
\textbf{\textrm{min}} \;\quad \sum_{e \in E} c_{e} y_{e} + \sum_{r \in \mathcal{R}} z_{r}\label{eq:LPgoal}\\\\[-2ex]
    \nonumber\\
\textbf{\textrm{s.t.}} \quad z_{r} + d_{r} \sum_{e \in s} y_{e} \geq v_{r} &\forall r \in \mathcal{R},\forall s \in S_{r}\label{eq:LPvalues}\\
    y_{e} \geq 0 \quad\quad\quad\quad\quad& \forall e \in E \label{eq:LPedges} \\
    z_{r} \geq 0 \quad\quad\quad\quad\quad& \forall r \in \mathcal{R} \label{eq:LPflows} \\[-6ex]
\nonumber
\end{eqnarray*}\\
\hline
\end{tabular}
\caption{\label{cap:UFPILPandDualLP} The integer linear program of
the unsplittable flow problem (top), and the dual of its
relaxation (bottom). Note that $S_{r}$ denotes the set of all the
simple paths between $s_{r}$ and $t_{r}$ in $G$, $S = \bigcup_{r
\in \mathcal{R}} S_{r}$, and $d_{s}$ and $v_{s}$ denote the
respective demand and value of path $s$, i.e.\ if $s \in S_{r}$
then $d_{s} = d_{r}$ and $v_{s} = v_{r}$.}
\end{figure}

Algorithm \textsf{Bounded-UFP}, formally described below, is a
primal-dual based algorithm for the $\Omega(\ln m)$-bounded
unsplittable flow problem. Informally, the algorithm maintains the
variables of the primal and dual programs, and in each iteration
selects to satisfy a request, which corresponds to the ``most
violated'' constraint of the dual linear program. Favorably, this
reduces to finding a (normalized) shortest path in the graph $G$,
whose edge weights correspond to the set of dual variables
$y_{e}$. It is worth noting that the algorithm, and part of its
analysis is in the spirit of the algorithm suggested by Briest et
al.\ \cite{BriestKV05}.

\begin{algorithm}
\caption{\textsf{Bounded-UFP}($\epsilon$) }
\begin{algorithmic}[1]
\Require An accuracy parameter $\epsilon \in (0,1]$ %
\Ensure A (request, path) pairs set $\mathcal{W}$, which holds the requests to be allocated \smallskip %
\State Let $\mathcal{L}$ be a list of all the requests, and let $\mathcal{W}$ be an empty set %
\State \textbf{for all} $r \in \mathcal{L}$ \textbf{do} $z_{r} = 0$ \textbf{end for} \label{alg:RequestInit} %
\State \textbf{for all} $s \in \mathcal{S}$ \textbf{do} $x_{s} = 0$ \textbf{end for} \label{alg:PathInit} %
\State \textbf{for all} $e \in E$ \textbf{do} $y_{e} = \frac{1}{c_{e}}$ \textbf{end for} \smallskip %
\While{$\big(\mathcal{L} \neq \emptyset$ and $\sum_{e \in E}c_{e} y_{e} \leq e^{\epsilon (B-1)}\big)$} \label{alg:StopCond}%
    \ForAll {$r \in \mathcal{L}$} \label{alg:PathSelStart}
        \State Let $p_{r}$ be the shortest path between $s_{r}$ and $t_{r}$ in $G$ with respect to the weights $y_{e}$, and%
        \Statex \quad\quad\quad let $|p_{r}| = \sum_{e \in p_{r}} y_{e}$ be its length %
    \EndFor \label{alg:PathSelEnd} %
    \State Let $\hat{r}$ be the request, which minimizes $\frac{d_{r}}{v_{r}}|p_{r}|$ with respect to every $r \in \mathcal{L}$ %
    \State \textbf{for all} $e \in p_{\hat{r}}$ \textbf{do} $y_{e} = y_{e} \cdot e^{\epsilon B d_{\hat{r}} / c_{e}}$ \textbf{end for} %
    \State Add $(\hat{r}, p_{\hat{r}})$ to $\mathcal{W}$, and remove $\hat{r}$ from $\mathcal{L}$ %
    \State Let $x_{p_{\hat{r}}} = 1$ and $z_{\hat{r}} = v_{\hat{r}}$ \label{alg:VarUpdate} %
\EndWhile \smallskip %
\State \textbf{return} $\mathcal{W}$ %
\end{algorithmic}
\end{algorithm}

\noindent We would like to note that since the path related
variables, and the request related variables, i.e.\ the $x_{s}$
and $z_{r}$ variables respectively, play no role in the execution
of the algorithm, lines \ref{alg:RequestInit}, \ref{alg:PathInit},
and \ref{alg:VarUpdate} are not regarded part of the algorithm.
Nevertheless, we decided not to neglect them from the algorithm's
description since they ease the analysis presentation.

\newpage

\subsection{Analysis} In this subsection, we will
prove the following theorem.

\begin{theorem} \label{th:UfpResult}
For any $\epsilon \in (0,1]$,
algorithm \textsf{Bounded-UFP}$(\frac{\epsilon}{6})$ returns a
feasible $(\scriptstyle(1 + \epsilon)$$\frac{e}{e-1})$-approximate
solution for the $\Omega(\frac{\ln m}{\epsilon^{2}})$-bounded
unsplittable flow problem,
runs in
polynomial-time, and is monotone and exact w.r.t.\ the demand and
value of every request.
\end{theorem}

\begin{corollary}
For all values $\epsilon \in (0,1]$,
there exists a polynomial-time truthful $(\scriptstyle(1 +
\epsilon)$$\frac{e}{e-1})$-approximation mechanism for the
$\Omega(\ln m)$-bounded unsplittable flow problem,
where every request's demand and value is
unknown.
\end{corollary}

\noindent We begin by introducing a notation that ease the
analysis presentation:
\begin{itemize}
\item Let $x_{s}^{i}$, $y_{e}^{i}$, and $z_{r}^{i}$ be the
respective values of the variables $x_{s}$, $y_{e}$, and $z_{r}$
at the end of the $i$-th iteration of the algorithm, where $i \geq
0$. Mind that we regard the end of iteration $0$ as the beginning
of the algorithm. Additionally, we let $(y^{i},z^{i})$ denote the
set of dual variables at the end of the $i$-th iteration.

\item Let $P(i) = \sum_{r \in \mathcal{R}} v_{r} \cdot (\sum_{s
\in S_{r}} x_{s}^{i})$ be the value of the primal solution at the
end of the $i$-th iteration, and let $P$ be the value of the
primal solution when the algorithm terminates. Notice that $P$ is
the sum of values of requests selected to be allocated by the
algorithm, i.e.\ the outcome of the algorithm. In addition, we let
$\Delta P(i) = P(i) - P(i-1)$ be the value in which the primal
solution is incremented in the $i$-th iteration.

\item Let $D_{1}(i) = \sum_{e \in E} c_{e} y_{e}^{i}$ and
$D_{2}(i) = \sum_{r \in \mathcal{R}} z_{r}^{i}$ be the respective
values of the first and second parts of the dual solution at the
end of the $i$-th iteration, and let $D(i) = D_{1}(i) + D_{2}(i)$.
Also, let $D$ denote the value of the optimal solution for the
dual linear program.

\item Let $\alpha(i)$ denote the normalized length of the path
selected after the end of the $i$-th iteration. Note that if path
$p$ is selected in the $(i+1)$-th iteration then $\alpha(i) =
\frac{d_{p}}{v_{p}} |p| = \frac{d_{p}}{v_{p}} \sum_{e \in p}
y_{e}^{i}$.
\end{itemize}

% CORRECTNESS %%%%%%%%%%%%%%%%%%%%%%%%%%%%%%%%%%%%%%%%%%%%%%%%%%%%%%
% TRUTHFULNESS %%%%%%%%%%%%%%%%%%%%%%%%%%%%%%%%%%%%%%%%%%%%%%%%%%%%%
\smallskip \noindent {\bf Correctness and Truthfulness.}
The following lemmas establish the feasibility of the solution, and the monotonicity
and exactness of the algorithm.

\begin{lemma} \label{th:UfpFeasibility}
Algorithm \textsf{Bounded-UFP}$(\epsilon)$ outputs a feasible solution.
\end{lemma}

\begin{proof}
Assume by contradiction that the output of the algorithm is not
feasible. Let $\bar{p}$ be the first path that induces a violation
in the capacity of edge $e$ in the $\ell$-th iteration, and let
$\mathcal{P}$ be the family of paths selected before the $\ell$-th
iteration, which consist of $e$. Since $\bar{p}$ induces a
capacity violation then $\sum_{p \in \mathcal{P}} d_{p} +
d_{\bar{p}} > c_{e}$. Clearly, since $d_{\bar{p}} \in (0,1]$ it
follows that $\sum_{p \in \mathcal{P}}d_{p} > c_{e} - 1$.
Consequently, we get that
\begin{eqnarray*}
c_{e} y_{e}^{\ell-1} = c_{e} y_{e}^{0} \prod_{p \in \mathcal{P}} e^{\frac{\epsilon B d_{p}}{c_{e}}} = e^{\frac{\epsilon B}{c_{e}} \sum_{p \in \mathcal{P}} d_{p}} \\
> e^{\epsilon B \frac{c_{e} - 1}{c_{e}}} \geq e^{\epsilon B \frac{B - 1}{B}} = e^{\epsilon (B-1)},
\end{eqnarray*}
where the last inequality results from the fact that
$\frac{x-1}{x}$ is an increasing monotonic function for all $x
\geq 1$, and since $c_{e} \geq B \geq 1$. Inspecting the main loop
stopping condition, i.e.\ line \ref{alg:StopCond} in the
algorithm, it follows that the algorithm had to exit the loop.
This implies that the algorithm could not have executed the
$\ell$-th iteration and thus, could not have selected $\bar{p}$, a
contradiction.~
\end{proof}

\begin{lemma} \label{th:UfpMonotone}
Algorithm \textsf{Bounded-UFP}$(\epsilon)$ is monotone and exact w.r.t. to
the demand and value of every request.
\end{lemma}

\begin{proof}
Consider a request $r$ selected to be routed using path $p_{r}$ in
the $\ell$-th iteration of the algorithm, which has a respective
demand and value of $d_{r}$ and $v_{r}$. Now, suppose that $r$ had
a demand of $\tilde{d_{r}} \leq d_{r}$, a value of $\tilde{v_{r}}
\geq v_{r}$, and the demands and values of all the other requests
were fixed. For the sake of monotonicity, we need to prove that
the algorithm would have selected $r$ in the latter case, i.e.\
when its demand and value were $\tilde{d_{r}}$ and
$\tilde{v_{r}}$, respectively. If $r$ is selected by the algorithm
in the first $\ell-1$ iterations then we are done. Otherwise, lets
consider the $\ell$-th iteration. One can easily observe that in
the first $\ell-1$ iterations of the algorithm, the same set of
requests is selected to be routed using the same set of paths
whether the demand and value of $r$ is $(d_{r}, v_{r})$ or
$(\tilde{d_{r}}, \tilde{v_{r}})$. Respectively, the same set of
unselected requests remain. Note that
$\frac{\tilde{d_{r}}}{\tilde{v_{r}}} \leq \frac{d_{r}}{v_{r}}$.
Hence, since the path $p_{r}$ minimizes $\frac{d_{p}}{v_{p}}
\sum_{e \in p} y_{e}$ over any path $p$, which corresponds to an
unselected request, when the demand and value of $r$ is $(d_{r},
v_{r})$, so it does when the demand and value of $r$ is
$(\tilde{d_{r}}, \tilde{v_{r}})$. This implies that $r$ must be
selected by the algorithm in the $\ell$-th iteration.

The exactness of the algorithm is clear, as the algorithm may
route the exact demand of every request selected, and may not
route anything otherwise.~
\end{proof}

% APPROXIMATION %%%%%%%%%%%%%%%%%%%%%%%%%%%%%%%%%%%%%%%%%%%%%%%%%%%%
\noindent {\bf Approximation.} We now turn to prove that the
algorithm yields an approximation ratio that approaches
$\frac{e}{e-1}$. We begin by stating three claims, which will be
utilized later.

\begin{claim} \label{th:AlgebraicClaim}
An increasing sequence $\{
\alpha_{0},\alpha_{1},\ldots,\alpha_{t+1},\alpha \}$
satisfies $
\sum_{i=0}^{t} \big(\frac{\alpha_{i+1}-\alpha_{i}}{\alpha -
\alpha_{i}}\big) \leq \ln \big(\frac{\alpha - \alpha_{0}}{\alpha -
\alpha_{t+1}}\big)$.
\end{claim}

\begin{proof}
For every $0 \leq i \leq t$,
$$
\frac{\alpha_{i+1}-\alpha_{i}}{\alpha - \alpha_{i}} \leq \ln
\Big(\frac{\alpha - \alpha_{i}}{\alpha - \alpha_{i+1}} \Big) =
\ln(\alpha - \alpha_{i}) - \ln(\alpha - \alpha_{i+1}) \ ,
$$
where the inequality follows from $\ln(1+x) \leq x$ by
substituting $x = \frac{\alpha_{i} - \alpha_{i+1}}{\alpha -
\alpha_{i}}$. Accordingly, this implies that
\begin{eqnarray*}
\sum_{i=0}^{t} \Big(\frac{\alpha_{i+1}-\alpha_{i}}{\alpha - \alpha_{i}}\Big) \leq \sum_{i=0}^{t} \big(\ln(\alpha - \alpha_{i}) - \ln(\alpha - \alpha_{i+1})\big) \\
= \ln(\alpha - \alpha_{0}) - \ln(\alpha - \alpha_{t+1}) = \ln \Big(\frac{\alpha - \alpha_{0}}{\alpha - \alpha_{t+1}}\Big) \ .
\end{eqnarray*}
\end{proof}

\begin{claim} \label{th:AlphaBound}
$\alpha(i) \leq \frac{D_{1}(i)}{D - D_{2}(i)}$, in every iteration
$i \geq 0$.
\end{claim}

\begin{proof}
Consider the $(i + 1)$-th iteration. Let $p$ denote the path that
is selected in this iteration. The path $p$ corresponds to an
unselected request such that $\frac{d_{p}}{v_{p}} \sum_{e \in
p}y_{e}^{i}$ is minimal. Namely, every other path $p'$, which
corresponds to another unselected request, satisfies
$$
\frac{d_{p'}}{v_{p'}} \sum_{e \in p'} y_{e}^{i} \geq
\frac{d_{p}}{v_{p}} \sum_{e \in p}y_{e}^{i} = \alpha(i) \text{, thus
} d_{p'} \sum_{e \in p'} \frac{y_{e}^{i}}{\alpha(i)}
\geq v_{p'} \ .
$$
This implies that if we multiply $y_{e}^{i}$ by $\alpha(i)^{-1}$,
for every $e \in E$, then all the dual linear program constraints
become satisfied. Consequently, the set of variables
$(y^{i}\alpha(i)^{-1},z^{i})$ constitutes a feasible fractional
solution to the dual linear program and therefore, $D \leq
D_{1}(i) \alpha(i)^{-1} + D_{2}(i)$.~
\end{proof}

\begin{claim} \label{th:D1DeltaBound}
$D_{1}(i + 1) \leq D_{1}(i) + B \epsilon (1 + \epsilon) \cdot
\Delta P(i + 1) \cdot \alpha(i)$, for every $i \geq 0$.
\end{claim}

\begin{proof}
Consider the $(i+1)$-th iteration. Let $p$ denote the path that is
selected in this iteration, and let $d_{p}$ and $v_{p}$ denote its
respective demand and value. Inspecting the algorithm, one can
derive that \setlength\arraycolsep{2pt}
\begin{eqnarray*}
\sum_{e \in E} c_{e} y_{e}^{i+1} & = & \sum_{\substack{e \in E \\ e \notin p}} c_{e} y_{e}^{i} + \sum_{e \in p} c_{e} y_{e}^{i} \cdot e^{\frac{\epsilon B d_{p}}{c_{e}}}\\
& \leq & \sum_{\substack{e \in E \\ e \notin p}} c_{e} y_{e}^{i} + \sum_{e \in p} c_{e} y_{e}^{i} \left(1 + \frac{\epsilon B d_{p}}{c_{e}} + \Big(\frac{\epsilon B d_{p}}{c_{e}}\Big)^{2}\right)\\
% & = & \sum_{\substack{e \in E}} c_{e} y_{e}^{i} + \sum_{e \in p} c_{e} y_{e}^{i} \cdot \left(\frac{\epsilon B d_{p}}{c_{e}} + \Big(\frac{\epsilon B d_{p}}{c_{e}}\Big)^{2}\right)\\
& \leq & \sum_{\substack{e \in E}} c_{e} y_{e}^{i} + B \epsilon (1 + \epsilon) d_{p} \sum_{e \in p} y_{e}^{i} \\
& = & \sum_{\substack{e \in E}} c_{e} y_{e}^{i} + B \epsilon (1 +
\epsilon) \cdot \Delta P(i+1) \cdot \alpha(i) \ .
\end{eqnarray*}
The first inequality is due to the fact that $e^{a} \leq 1 + a +
a^{2}$ for any $a \in [0,1]$, and the fact that $\frac{\epsilon B
d_{p}}{c_{e}} \in (0,1]$. The second inequality holds since
\begin{equation}
c_{e} \cdot \left(\frac{\epsilon B d_{p}}{c_{e}} +
\Big(\frac{\epsilon B d_{p}}{c_{e}}\Big)^{2}\right) \leq \epsilon
B d_{p} + \epsilon^{2} B d_{p} =  B \epsilon (1 + \epsilon) d_{p}
\ , \label{eq:D1Bound}
\end{equation}
where the inequality in (\ref{eq:D1Bound}) follows from the
observations that $d_{p}^{2} \leq d_{p}$, and $\frac{B}{c_{e}}
\leq 1$. Finally, the last equality follows from the definition of
$\alpha(i)$, which can be rewritten as $d_{p} \sum_{e \in p}
y_{e}^{i} = v_{p} \alpha(i)$, and the observation that $v_{p}$ is
the value in which the primal solution is incremented in the
$(i+1)$-th iteration. Recalling that $D_{1}(i) = \sum_{e \in E}
c_{e} y_{e}^{i}$ completes the proof.~
\end{proof}

\noindent We are now ready to establish the approximation
guarantee of the algorithm.

\begin{lemma} \label{th:UfpApprox}
Algorithm \textsf{Bounded-UFP}$(\epsilon)$ returns an
$(\scriptstyle(1 + 6\epsilon)$$\frac{e}{e-1})$-approximate
solution for the $\frac{\ln m}{\epsilon^{2}}$-bounded unsplittable
flow problem, for any $\epsilon \in (0,\frac{1}{6}]$.
\end{lemma}
\begin{proof}
One can easily notice, by inspecting the stoping condition of the
main loop, that when the algorithm terminates, either $\mathcal{L}
= \emptyset$ or $\sum_{e \in E} c_{e} y_{e} > e^{\epsilon (B-1)}$.
If $\mathcal{L} = \emptyset$ then it follows that the algorithm
succeeded to satisfy all the requests and thus, its output is
optimal. Consequently, in the remainder of the proof, we shall
assume that $\sum_{e \in E} c_{e} y_{e} > e^{\epsilon (B-1)}$. For
every $i \geq 0$, \setlength\arraycolsep{2pt}
\begin{eqnarray*}
D_{1}(i + 1) & \leq & D_{1}(i) + B \epsilon (1 + \epsilon) \cdot \Delta P(i + 1) \cdot \alpha(i) \\
& \leq & D_{1}(i) + B \epsilon (1 + \epsilon) \cdot \Delta P(i+1) \cdot \frac{D_{1}(i)}{D - D_{2}(i)} \\
& = & D_{1}(i) \Big(1 + B \epsilon (1 + \epsilon) \frac{\Delta  P(i+1)}{D - D_{2}(i)}\Big) \\
& \leq & D_{1}(i) e^{\big(B \epsilon (1 + \epsilon) \frac{\Delta
P(i+1)}{D - D_{2}(i)}\big)} \ ,
\end{eqnarray*}
where the first and second inequalities follow from
Claim~\ref{th:D1DeltaBound} and Claim~\ref{th:AlphaBound},
respectively, and the last inequality is due to the fact that $1 +
a \leq e^{a}$. This implies that
\begin{eqnarray*}
D_{1}(i + 1) & \leq & D_{1}(0) e^{\big(B \epsilon (1 + \epsilon) \sum_{j=0}^{i} \frac{\Delta P(j + 1)}{D - D_{2}(j)}\big)} \\
& \leq & e^{\big(B  \epsilon^{2} + B \epsilon (1 + \epsilon) \sum_{j=0}^{i} \frac{\Delta P(j + 1)}{D - D_{2}(j)}\big)} \ ,
\end{eqnarray*}
where the first inequality results from the expansion of the
recursion, and the second one follows from $D_{1}(0) \leq
e^{B \epsilon^{2}}$, which is obtained by noticing that $D_{1}(0)
= \sum_{e \in E} c_{e} y_{e}^{0} = m$, and recalling that $B \geq
\frac{\ln m}{\epsilon^{2}}$ by definition. Lets assume that the
algorithm terminates after $t + 1$ iterations. Accordingly, using
our prior assumption that $\sum_{e \in E} c_{e} y_{e} >
e^{\epsilon (B-1)}$, we get that
$$
e^{\epsilon (B-1)} < D_{1}(t + 1) \leq e^{\big(B  \epsilon^{2} + B
\epsilon (1 + \epsilon) \sum_{j=0}^{t} \frac{\Delta P(j + 1)}{D -
D_{2}(j)}\big)} \ .
$$
One can validate that $D_{2}(j) = P(j)$ by inspecting the
variables alterations in line \ref{alg:VarUpdate} of the
algorithm, and their affect on $D_{2}(j)$ and $P(j)$. Hence, we
derive that $\frac{\Delta P(j + 1)}{D - D_{2}(j)} = \frac{P(j + 1)
- P(j)}{D - P(j)}$. Consequently, we can apply
Claim~\ref{th:AlgebraicClaim}, while recalling that $P(0) = 0$ and
$P(t+1) = P$, and yield
$$
\textstyle \epsilon (B-1) < B \epsilon^{2} + B \epsilon (1 +
\epsilon) \ln \big(\frac{D}{D - P}\big) \ .
$$
Because $\frac{1 - 2\epsilon}{1 + \epsilon} \leq \frac{\epsilon
(B-1) - B \epsilon^{2}}{B \epsilon (1 + \epsilon)}$, and since $1
- 3 \epsilon \leq \frac{1 - 2\epsilon}{1 + \epsilon}$ for any
positive $\epsilon$, we attain $1 - 3\epsilon \leq \ln
\big(\frac{D}{D - P}\big)$. This can be simplified further to give
$\frac{D}{P} \leq \frac{1}{e^{(1 - 3\epsilon)} - 1} + 1 \leq
{\scriptstyle (1 + 6\epsilon)} \frac{e}{e-1}$. Recall that $D$ is
the value of the optimal solution for the dual linear program and
thus, using the weak LP duality completes the proof.~
\end{proof}

\begin{exrefproof}{Theorem}{\ref{th:UfpResult}}
The correctness of the algorithm is due to
Lemma~\ref{th:UfpFeasibility}, the approximation guarantee is
established in Lemma~\ref{th:UfpApprox}, and the monotonicity and
exactness are presented in Lemma~\ref{th:UfpMonotone}. Finally, it
is clear that the running time of the algorithm is polynomial, and
in fact, if we denote the number of requests by $|\mathcal{R}|$,
one can easily validate that the number of iterations is bounded
by $|\mathcal{R}|$, and every iteration takes time proportional to
$|\mathcal{R}|$ shortest path computations.~
\end{exrefproof}

\subsection{Inapproximability result} \label{cha:UfpInapprox}
In the following, we introduce two input instances for the problem
under consideration that lower bound the performance guarantee of
any algorithm, which is part of the reasonable iterative path
minimizing algorithms family. Specifically, the first input
instance proves that any such algorithm cannot yield an
approximation guarantee better than $\frac{e}{e-1} - o(1)$. This
demonstrates that the analysis of algorithm \textsf{Bounded-UFP}
is tight, and that it is the ``best'' algorithm in the
aforementioned family of algorithms. The second input instance
establishes a lower bound of $\frac{4}{3}$ on the approximation
ratio of any such algorithm for this problem in its utmost
generality, i.e.\ when the underlying graph is undirected and the
minimal edge capacity is arbitrarily large. In particular, this
suggests that even if we ease the problem setting, e.g.\ assume
that the minimal capacity of an edge is $\Omega(m)$ instead of
$\Omega(\ln m)$, no algorithm in the aforesaid family can achieve
PTAS.

Prior to describing the finer details of our approach, we
introduce the notion of a reasonable function, which is a key
ingredient in the definition of a reasonable iterative path
minimizing algorithm. Note that reasonable functions have a
similar flavor to the min functions introduced by Archer and
Tardos \cite{ArcherT02}. Nevertheless, they are still quite
different. Let $S = \bigcup_{r \in \mathcal{R}} \{ p : p \text{ is
a simple path between } s_{r} \text{ and } t_{r} \}$.

\begin{definition} \label{def:Reasonable}
Let $g: S \rightarrow \bbR$ be a function, which assigns an
arbitrary priority to every path. Such a function is called
\emph{reasonable} if under the assumption that the capacities of
all the edges are identical, and both the demand and value of
every request are unit, it follows that $g(q) \leq g(q')$ for any
valid unsplittable flow, and any two paths $q,q' \in S$ that
satisfy
\begin{itemize}
\item $q$ consists of $k$ edges, $q'$ consists of $k'$ edges, and
$k \leq k'$.

\item $f_{i} \leq f_{i}'$, for every $1 \leq i \leq k$, where
$(f_{1}, f_{2},\ldots, f_{k})$ and $(f_{1}', f_{2}',\ldots,
f_{k'}')$ are non-increasing vectors, which indicate the flow
routed through the edges of $q$ and $q'$ with respect to the valid
unsplittable flow.
\end{itemize}
\end{definition}

\begin{definition}
An algorithm is referred to as \emph{reasonable iterative path
minimizing algorithm}, if it iteratively selects a path that
minimizes a reasonable function over all the paths that correspond
to unselected requests.
\end{definition}

One can verify that algorithm \textsf{Bounded-UFP} minimizes the
function $h(p) = \frac{d_{p}}{v_{p}} \sum_{e \in p}
\frac{1}{c_{e}}e^{\frac{\epsilon B f_{e}}{c_{e}}}$, where $f_{e}$
denotes the flow routed through edge $e$. This function is
reasonable since when we assume that both the demand and value of
every request are unit, and the capacities of all the edges are
identical, say $B$, then it reduces to $\tilde{h}(p) = \frac{1}{B}
\sum_{e \in p} e^{\epsilon f_{e}}$, which clearly satisfies
$\tilde{h}(q) \leq \tilde{h}(q')$, for any two paths $q$ and $q'$
that meet the properties indicated in
Definition~\ref{def:Reasonable}. Consequently, algorithm
\textsf{Bounded-UFP} is a reasonable iterative path minimizing
algorithm. We note that reasonability captures a broad class of
functions. For example, the reasonable function $h_{1}(p) = \ln (1
+ |p|) \cdot h(p)$ is similar to the function used by algorithm
\textsf{Bounded-UFP} but it is mildly biased towards paths with
less edges. Another example of a reasonable function is $h_{2}(p)
= \frac{d_{p}}{v_{p}} \prod_{e \in p} \frac{f_{e}}{c_{e}}$
although it is not clear why anyone would like to use it. We are
now ready to establish the main result of this subsection.

\begin{theorem}
The approximation ratio of any reasonable iterative path
minimizing algorithm for the unsplittable flow problem when the
graph is directed cannot be better than $\frac{e}{e-1} - o(1)$,
for $B = o(m^{1/2})$.
\end{theorem}
\begin{proof}
\begin{figure}[!hbt]
\centerline{ \scalebox{0.60}{ \input{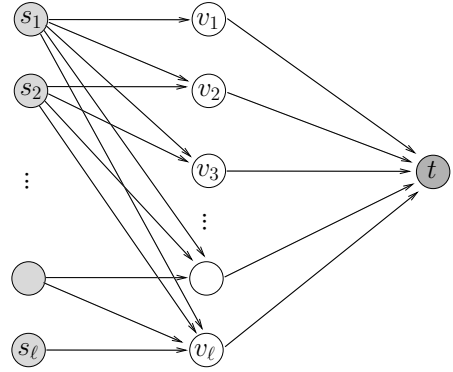} } }
\caption{A directed graph in which every vertex $s_{i}$ has a
directed edge to every vertex $v_{j}$ such that $j \geq i$.
Additionally, the capacities of all the edges are identical and
equal to $B$.} \label{cap:directedLowerBound}
\end{figure}
Suppose we are given the directed graph $G=(V,E)$ schematically
described in Figure~\ref{cap:directedLowerBound}, and the set of
requests is
$$
\mathcal{R} = \{ \underbrace{(s_{1},t,1,1)}_{B \text{ requests }},
\underbrace{(s_{2},t,1,1)}_{B \text{ requests }}, \ldots,
\underbrace{(s_{\ell},t,1,1)}_{B \text{ requests }} \} \ .
$$
In order to simplify the presentation and analysis of the lower
bound instance, we introduce the following assumption, which will
be tackled later. We assume that when there is more than one path,
which minimizes the value of the reasonable function used by the
iterative path minimizing algorithm, the algorithm selects one of
them arbitrarily. Accordingly, we premise that it selects a path
$(s_{i},v_{j},t)$ in which $i$ is \emph{minimal}, and $j$ is
\emph{maximal} with respect to all the minimizing paths that their
source vertex is $s_{i}$. For example, in the initial $B$
iterations of the algorithm, all the requests that their terminal
vertices are $(s_{1},t)$ are satisfied using paths that use
vertices $v_{j}$ such that $j = \ell, \ldots ,\ell - B + 1$. In
the subsequent $B$ iterations, all the requests that their
terminal vertices are $(s_{2},t)$ are satisfied using paths that
use vertices $v_{j}$ such that $j = \ell - B, \ldots, \ell - 2B +
1$, and so on. Simulating the execution of a reasonable iterative
path minimizing algorithm, while ignoring integrality issues that
will be resolved later, we get that
\begin{itemize}
\item For any integer $1 \leq q \leq B$, at the end of the first
$\ell \sum_{r=1}^{q}(\frac{B}{B+1})^{r}$ iterations, all and only
the requests, whose terminal vertices are $(s_{i},t)$ such that $i
\leq \ell \cdot (1 - (\frac{B}{B+1})^{q})$, are satisfied, and all
the $(v_{j},t)$ edges such that $j > \ell \cdot (1 -
(\frac{B}{B+1})^{q})$ have a flow load of $q$ or equivalently, a
residual capacity of $B - q$.

\item After $\ell \sum_{r=1}^{B}(\frac{B}{B+1})^{r}$ iterations,
the algorithm cannot route more requests and thus, it
stops\footnote{An algorithm might stop even sooner, e.g.\
algorithm \textsf{Bounded-UFP} stops after the while condition
fails. However, analyzing the case that the algorithm stops when
it cannot route more requests just affirms the lower bound.}.
\end{itemize}
Consequently, since all and only the requests, whose terminal
vertices are $(s_{i},t)$ such that $i \leq \ell \cdot (1 -
(\frac{B}{B+1})^{B})$, are satisfied when the algorithm stops, it
follows that the value of the solution that the algorithm outputs
is at most $ B \ell \cdot (1 - (\frac{B}{B+1})^{B}) = B \ell \cdot
(1 - (1 - \frac{1}{B+1})^{B}) \leq B \ell \cdot (1 -
\frac{1}{e})$. On the other hand, an optimal solution clearly has
a value of $B \ell$, e.g.\ route every request of the form
$(s_{i},t,1,1)$ through the directed path $(s_{i}, v_{i}, t)$.
Thus, we get that the approximation ratio of the algorithm cannot
be better than $\frac{e}{e - 1}$.

We now drop the integrality assumption. Namely, in
the above analysis, we have assumed that $\ell
\sum_{r=1}^{q}(\frac{B}{B+1})^{r}$ is integral, for any integer
$q$. This clearly may not be true. However, one can resolve this
issue by applying a more careful analysis, and yield that only the
requests, whose terminal vertices are $(s_{i},t)$ such that $i
\leq \ell \cdot (1 - (\frac{B}{B+1})^{q}) +
\sum_{k=0}^{q-1}(\frac{B}{B+1})^k$, may become satisfied. Since
$\sum_{k=0}^{B-1}(\frac{B}{B+1})^k \leq B$, it follows that the
value of the solution that the algorithm achieves might increase
by no more than $B^{2}$, in respect to $B \ell \cdot (1 -
\frac{1}{e})$. Since the number of edges in the graph is $m = \ell
+ \sum_{k=1}^{\ell}k \leq 2 \ell^{2}$, and $B = o(m^{1/2})$, we
obtain that $B = o(\ell)$. Consequently, the analysis of the lower
bound degrades by at most $o(1)$, i.e.\ the approximation ratio of
any algorithm cannot be better than $\frac{e}{e - 1} - o(1)$.

Next, we tackle the decisions assumption. Specifically, we have
assumed that the decision of any algorithm between same valued
minimizing paths is arbitrary and hence, one may ask if a specific
tie-breaking rule can lead to better results. We can resolve this
matter by constructing a similar input instance, which forces any
algorithm to make similar ``bad'' decisions. Essentially, one way
to achieve it is to replace every $(s_{i},v_{j})$ edge by a
directed path with $i \ell + 1 - j$ edges. The reason that an
algorithm makes ``bad'' decisions on this instance dues to the
reasonability property, i.e.\ any reasonable algorithm ``prefers''
paths with less edges. Note that this instance supports the same
lower bound, but has a somewhat stricter constraint on the value
of $B$, i.e.\ since $m = O(\ell ^4)$, $B$ needs to satisfy $B =
o(m^{1/4})$.~
\end{proof}

The next theorem
demonstrates that even if we
ease the problem setting, no reasonable iterative path minimizing
algorithm can achieve PTAS.

\begin{theorem} \label{th:LowerBoundUfp}
The approximation ratio of any reasonable iterative path
minimizing algorithm for the unsplittable flow problem cannot be
better than $\frac{4}{3}$, for any B, and even when the graph is
undirected.
\end{theorem}

\begin{proof}
\begin{figure}[!hbt]
\centerline{\scalebox{0.60}{\input{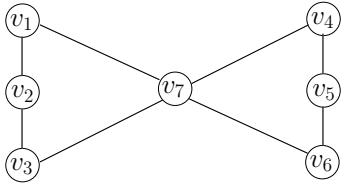}}}
\caption{An undirected graph in which the capacities of all the
edges are identically equal to $B$.}
\label{cap:undirectedLowerBound}
\end{figure}
Suppose we are given the undirected graph $G=(V,E)$ schematically
described in Figure~\ref{cap:undirectedLowerBound}, and the set of
requests is
$$
\mathcal{R} = \{ \underbrace{(v_{1},v_{3},1,1)}_{B \text{ requests
}}, \underbrace{(v_{4},v_{6},1,1)}_{B \text{ requests }},
\underbrace{(v_{1},v_{6},1,1)}_{B \text{ requests }},
\underbrace{(v_{3},v_{4},1,1)}_{B \text{ requests }} \} \ .
$$
Clearly, an optimal solution for this instance has a value of
$4B$, e.g.\ route every request of the form $(v_{1},v_{3},1,1)$
through the path $(v_{1}, v_{2}, v_{3})$, any request of the
form $(v_{4},v_{6},1,1)$ through the path $(v_{4}, v_{5}, v_{6})$,
all the requests of the form $(v_{1},v_{6},1,1)$ through the path
$(v_{1}, v_{7}, v_{6})$, and every request of the form
$(v_{3},v_{4},1,1)$ through the path $(v_{3}, v_{7}, v_{4})$.

Simulating the execution of a reasonable iterative path minimizing
algorithm in the initial four iterations, one can easily validate
that the algorithm may select the four paths $(v_{1}, v_{7},
v_{3})$,$(v_{4}, v_{7}, v_{6})$, $(v_{1}, v_{2}, v_{3})$, and
$(v_{4}, v_{5}, v_{6})$ since each one of these paths is one of
the minimizing paths in the corresponding iteration, and by that
satisfy two $(v_{1}, v_{3}, 1, 1)$ requests and two $(v_{4},
v_{6}, 1, 1)$ requests. In addition, notice that at the end of
this four iterations phase, every edge has a residual capacity of
$B - 1$. Arguments similar to those used in this initial four
iterations phase can be applied in another $\frac{B}{2} - 1$
phases, each of four iterations, to demonstrate that the algorithm
acts exactly the same. Consequently, after $\frac{B}{2}$ phases,
every edge has a residual capacity of $\frac{B}{2}$, all the
$(v_{1}, v_{3}, 1, 1)$ requests and the $(v_{4}, v_{6}, 1, 1)$
requests were satisfied, and the remaining requests are
$$
\bar{\mathcal{R}} = \{ \underbrace{(v_{1},v_{6},1,1)}_{B \text{
requests }}, \underbrace{(v_{3},v_{4},1,1)}_{B \text{ requests }}
\} \ .
$$
In this current state, any algorithm can satisfy at most $B$
requests from $\bar{\mathcal{R}}$. This is the result of the fact
that any path from $v_{1}$ to $v_{6}$ and any path from $v_{3}$ to
$v_{4}$ must use either edge ($v_{1},v_{7}$) or edge
($v_{3},v_{7}$), and the fact that the total residual capacity of
these edges sums to $B$. Thus, the solution that the algorithm
outputs has value of at most $3B$.~
\end{proof}

\begin{corollary}
No reasonable iterative path minimizing algorithm for the
unsplittable flow problem can yield a PTAS.
\end{corollary}

% SINGLE MINDED MULTI UNIT COMBINATORIAL AUCTION PROBLEM %%%%%%%%%%%
\section{Single-minded Multi-unit \\ Combinatorial Auction} \label{cha:MucaSect}
\subsection{The algorithm}
In this subsection, we design a deterministic monotone algorithm
for the $\Omega(\ln m)$-bounded multi-unit combinatorial auction
problem, whose approximation ratio approaches $\frac{e}{e-1}$.
We begin by demonstrating that the single-minded
multi-unit combinatorial auction problem can be formulated as a
simplified special case of the integer linear program of the
unsplittable flow problem, and then we turn to specialize
algorithm \textsf{Bounded-UFP} for the problem under
consideration.

The single-minded multi-unit combinatorial auction problem can be
formulated as a special case of the integer linear program of the
unsplittable flow problem by letting $S_{r}$ to denote the
singleton set of $U_{r}$, i.e.\ $S_{r}= \{ U_{r} \}$, and
replacing $e$, $E$ and $d_{s}$ in the integer linear program of
the unsplittable flow problem with $u$, $U$ and $1$, respectively.
Similarly, the relaxation of the integer linear program of the
single-minded multi-unit combinatorial auction problem, and its
dual can also be formulated as special cases of the corresponding
linear programs. Consequently, the primal-dual algorithm
\textsf{Bounded-MUCA}, formally described below, is a specialized
version of algorithm \textsf{Bounded-UFP}, in which the path
selection procedure, i.e.\ lines
\ref{alg:PathSelStart}-\ref{alg:PathSelEnd} in algorithm
\textsf{Bounded-UFP}, was neglected, and the demand terms were
omitted.

\begin{algorithm}
\caption{\textsf{Bounded-MUCA}($\epsilon$) } \label{cap:BMucaAlg}
\begin{algorithmic}[1]
\Require An accuracy parameter $\epsilon \in (0,1]$ %
\Ensure A set $\mathcal{W}$ of requests to be satisfied \smallskip %
\State Let $\mathcal{L}$ be a list of all the requests, and let $\mathcal{W}$ be an empty set %
\State \textbf{for all} $u \in U$ \textbf{do} $y_{u} = \frac{1}{c_{u}}$ \textbf{end for} \smallskip %
\While{$\big(\mathcal{L} \neq \emptyset$ and $\sum_{u \in U}c_{u} y_{u} \leq e^{\epsilon (B-1)}\big)$} %
    \State Let $\hat{r}$ be the request, which minimizes $\frac{1}{v_{r}}\sum_{u \in U_{r}} y_{u}$ with respect to every $r \in \mathcal{L}$ %
    \State \textbf{for all} $u \in U_{\hat{r}}$ \textbf{do} $y_{u} = y_{u} \cdot e^{\epsilon B / c_{u}}$ \textbf{end for} %
    \State Add $\hat{r}$ to $\mathcal{W}$, and remove $\hat{r}$ from $\mathcal{L}$ %
\EndWhile \smallskip %
\State \textbf{return} $\mathcal{W}$ %
\end{algorithmic}
\end{algorithm}

\begin{theorem} \label{th:MucaResult}
The algorithm \textsf{Bounded-MUCA}$(\frac{\epsilon}{6})$ returns a
feasible $(\scriptstyle(1 + \epsilon)$$\frac{e}{e-1})$-approximate
solution for the $\Omega(\frac{\ln m}{\epsilon^{2}})$-bounded
multi-unit combinatorial auction problem, for any $\epsilon \in
(0,1]$, runs in polynomial-time, and is monotone and exact w.r.t.\
the value of each request.
\end{theorem}
\begin{proof}
Since algorithm \textsf{Bounded-MUCA} is a simplified version of
algorithm \textsf{Bounded-UFP}, the analysis of
Theorem~\ref{th:UfpResult} also applies in this case.~
\end{proof}

It is worth noting that the algorithm can even be employed to a
generalized version of the problem in which the demand of every
request, i.e.\ the desired bundle of items, is part of the type of
the request and therefore, agents may be dishonest about it. Note
that in this case, monotonicity can be easily established by
arguments similar to those used in Theorem~\ref{th:UfpMonotone},
and the additional observation that $\sum_{u \in \tilde{{U}_{r}}}
y_{u} \leq \sum_{u \in U_{r}} y_{u}$, for any $\tilde{{U}_{r}}
\subseteq U_{r}$. Remark that this setting is referred to as the
\emph{unknown} single-minded case \cite{MualemN02}.

\begin{corollary}
For all values $\epsilon \in (0,1]$,
there exists a polynomial-time truthful $(\scriptstyle(1 +
\epsilon)$$\frac{e}{e-1})$-approximation mechanism for the
$\Omega(\ln m)$-bounded multi-unit combinatorial auction problem
among unknown single-minded agents.
\end{corollary}

\subsection{Inapproximability result} \label{cha:MucaInapprox}
We demonstrate that any reasonable iterative bundle minimizing
algorithm cannot achieve an approximation ratio better than
$\frac{4}{3}$. Essentially, this suggests that no algorithm in the
aforesaid family can achieve PTAS.

We start by formally defining the notion of a reasonable iterative
bundle minimizing algorithm. We remark that the following
definitions are just a refinement of the definitions made for the
unsplittable flow problem. Let $S = \bigcup_{r \in \mathcal{R}} \{
U_{r} \}$.

\begin{definition} \label{def:ReasonableMuca}
Let $g: S \rightarrow \bbR$ be a function, which assigns an
arbitrary priority to every bundle. Such a function is called
\emph{reasonable} if under the assumption that the multiplicities
of all the items are identical, and the value of every request is
unit, it follows that $g(T) \leq g(T')$ for any valid multi-unit
allocation\footnote{A \emph{valid multi-unit allocation} can be
succinctly described as an allocation of non-identical items to
requests such that the number of allocated copies of every item
does not exceed its multiplicity.}, and any two bundles $T,T' \in
S$ that satisfy
\begin{itemize}
\item $T$ consists of $k$ items, $T'$ consists of $k'$ items, and
$k \leq k'$.

\item $f_{i} \leq f_{i}'$, for every $1 \leq i \leq k$, where
$(f_{1}, f_{2},\ldots, f_{k})$ and $(f_{1}', f_{2}',\ldots,
f_{k'}')$ are non-increasing vectors, which indicate the number of
allocated copies of the items of $T$ and $T'$ with respect to the
valid multi-unit allocation.
\end{itemize}
\end{definition}

\begin{definition}
An algorithm is referred to as \emph{reasonable iterative bundle
minimizing algorithm}, if it iteratively selects a bundle that
minimizes a reasonable function over all the bundles that
correspond to unselected requests.
\end{definition}

Note that algorithm \textsf{Bounded-MUCA} minimizes the function
$h(s) = \frac{1}{v_{s}} \sum_{u \in s} \frac{1}{c_{u}}
e^{\frac{\epsilon B f_{u}}{c_{u}}}$, where $f_{u}$ denotes the
number of allocated copies of item $u$. One can easily argue that
this function is reasonable and thus, algorithm
\textsf{Bounded-MUCA} is a reasonable iterative bundle minimizing
algorithm. We are now ready to prove the core result of this
subsection.

\begin{theorem}\label{th:LowerBoundMUCA}
The approximation ratio of any reasonable iterative bundle
minimizing algorithm for the single-minded multi-unit
combinatorial auction problem cannot be better than $\frac{4}{3}$.
\end{theorem}

\begin{proof}
Let $m$ be a multiple of $p \cdot (p+1)$, where $p \geq 3$ is a
constant odd integer. Suppose we are given a set $U$ of $m$ items
such that the multiplicities of all the items are identical and
equal to $B$, and let $\bigcup_{i=1}^{p} \bigcup_{j=1}^{p+1}
U_{i,j}$ be a partition of $U$ into $p \cdot (p+1)$ disjoint sets,
each consists of $\frac{m}{p \cdot (p+1)}$ items. Additionally,
suppose that $\mathcal{R}$ consists of unit value requests of two
types:
\begin{enumerate}
\item $\frac{B}{2}$ requests that consist of the items $U_{\ell} =
\bigcup_{j=1}^{p+1} U_{\ell, j}$, for every $\ell=1,\ldots,p$.
\item $\frac{B}{2}$ requests that consist of the items
$U_{1,2\ell-1} \cup U_{1,2\ell} \cup \bigcup_{i=2}^{p}
U_{i,2\ell-1}$, for every $\ell= 1, \ldots, \frac{p+1}{2}$, and
$\frac{B}{2}$ requests that consist of the items $U_{1,2\ell-1}
\cup U_{1,2\ell} \cup \bigcup_{i=2}^{p} U_{i,2\ell}$, for every
$\ell= 1, \ldots, \frac{p+1}{2}$.
\end{enumerate}
Figure \ref{cap:mucaLowerBound} schematically describes a concrete
input instance.
\begin{figure}[!hbt]
\centerline{ \scalebox{0.70}{ \input{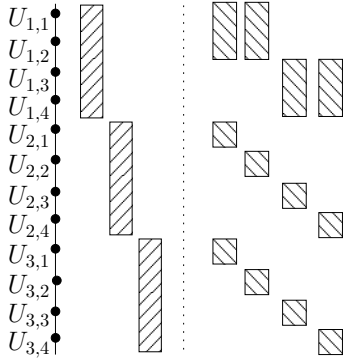} } }
\caption{The input instance in the case of $p = 3$, and $m = 12$.
Note that the set of items $\bigcup_{i=1}^{3} \bigcup_{j=1}^{4}
U_{i,j}$ are represented by the black dots on the left of the
figure, and every collection of $\frac{B}{2}$ requests is
represented by a vertical group of rectangles such that each
rectangle consists of the set of items, whose representing dots
are in the left-projection zone of the rectangle. Also note that
the first type of requests are located left of the dotted line,
whereas the second type of requests appear on its right.}
\label{cap:mucaLowerBound}
\end{figure}

One can verify that the an optimal solution for this instance has
a value of $p B$, for example by selecting all the requests except
for the $\frac{B}{2}$ requests that consist of $U_{1}$.

Simulating the execution of a reasonable iterative bundle
minimizing algorithm, one can easily validate that the algorithm
may incrementally select all the requests of the first type, e.g.
it may repeatedly select a request that consists of $U_{1}$, then
a request that consists of $U_{2}$ and so on until $U_{p}$.
Consequently, after $\frac{B}{2}$ phases, each of $p$ steps, all
the requests of the first type are satisfied, the value of the
current solution is $\frac{p}{2} B$, and every item has a residual
multiplicity of $\frac{B}{2}$. Notice that $U_{1}$ consists of
$\frac{m}{p}$ items, and any request of the second type consists
of $\frac{2m}{p \cdot (p+1)}$ items of $U_{1}$. Hence, by simple
counting arguments it follows that in the current state, any
algorithm cannot satisfy more than $(\frac{B}{2} \cdot \frac{m}{p}
) / \frac{2m}{p \cdot (p+1)} = \frac{p+1}{4} B$ requests of the
second type. Therefore, we obtain that the value of the solution
that any reasonable iterative bundle minimizing algorithm outputs
is no more than $\frac{3p+1}{4} B$ and thus, as $p$ tends to
infinity, the inapproximability ratio approaches $\frac{4}{3}$.~
\end{proof}

\begin{corollary}
No reasonable iterative bundle minimizing algorithm for the
single-minded multi-unit combinatorial auction problem yields a
PTAS.
\end{corollary}

\section{Unsplittable Flow with \\ Repetitions Problem} \label{cha:UfpRepeatSect}
In this section, we study the $\Omega(\ln m)$-bounded unsplittable
flow with repetitions problem. This problem is a variant of the
corresponding unsplittable flow problem in which we are allowed to
satisfy every request multiple times using possibly multiple
paths, and the profit gained is proportional to the number of
times that every request is satisfied. In sharp contrast with our
prior results, we demonstrate that this version admits a
deterministic primal-dual based algorithm, which yields an $(1 +
\epsilon)$-approximation.

\subsection{The algorithm}
In the following, we contrive an $(1 + \epsilon)$-approximation
algorithm, named \textsf{Bounded-UFP-Repeat}, for the $\Omega(\ln
m)$-bounded unsplittable flow with repetitions problem, which is
based on a primal-dual approach. The following theorem
digests the
properties of the algorithm.

\begin{theorem} \label{th:UfpRepeatResult}
Algorithm \textsf{Bounded-UFP-Repeat}$(\frac{\epsilon}{6})$ is an $(1 +
\epsilon)$-approximation for the $\Omega(\frac{\ln
m}{\epsilon^{2}})$-bounded unsplittable flow with repetitions
problem, for any $\epsilon \in (0,1]$, with running time
polynomial in $m$ and $\frac{\max_{e} \{ c_{e} \}}{\min_{r} \{
d_{r} \}}$.
\end{theorem}

\begin{algorithm}
\caption{\textsf{Bounded-UFP-Repeat}($\epsilon$) }
\begin{algorithmic}[1]
\Require An accuracy parameter $\epsilon \in (0,1]$ %
\Ensure A (request, path) pairs multiset $\mathcal{W}$, which holds the requests to be allocated \smallskip %
\State Let $\mathcal{L}$ be a list of all the requests, and let $\mathcal{W}$ be an empty multiset %
\State \textbf{for all} $e \in E$ \textbf{do} $y_{e} = \frac{1}{c_{e}}$ \textbf{end for} \smallskip %
\While{$\big(\sum_{e \in E}c_{e} y_{e} \leq e^{\epsilon (B-1)}\big)$} %
    \For {\textbf{every} $r \in \mathcal{L}$} %
        \State Let $p_{r}$ be the shortest path between $s_{r}$ and $t_{r}$ in $G$ with respect to the weights $y_{e}$, and%
        \Statex \quad\quad\quad let $|p_{r}| = \sum_{e \in p_{r}} y_{e}$ be its length %
    \EndFor  %
    \State Let $\hat{r}$ be the request, which minimizes $\frac{d_{r}}{v_{r}}|p_{r}|$ with respect to every $r \in \mathcal{L}$ %
    \State \textbf{for all} $e \in p_{\hat{r}}$ \textbf{do} $y_{e} = y_{e} \cdot e^{\epsilon B d_{\hat{r}} / c_{e}}$ \textbf{end for} %
    \State Add $(\hat{r}, p_{\hat{r}})$ to $\mathcal{W}$ %
\EndWhile \smallskip %
\State \textbf{return} $\mathcal{W}$ %
\end{algorithmic}
\end{algorithm}

In
Figure~\ref{cap:UFPRepeatILPandDualLp}, we present the primal-dual
formulation of the underlying problem.

\begin{figure}[!hbt]
\centering
\begin{tabular}{|p{6.5cm}|}
\hline
\begin{eqnarray*}
\nonumber \\[-6ex]
\textbf{\textrm{max}} \;\quad \sum_{r \in \mathcal{R}} v_{r} \cdot \Big(\sum_{s \in S_{r}} x_{s}\Big)\label{eq:ILPgoalR}\\\\[-2ex]
    \nonumber\\
\textbf{\textrm{s.t.}} \quad \sum_{s \in S | e \in s} x_{s}   d_{s} \leq c_{e}\quad&  \forall e \in E\label{eq:ILPedgesR}\\
    x_{s} \in \bbN \quad\quad\quad\quad\quad& \forall s \in S \label{eq:ILPvarsR}
\\[-7ex] \nonumber
\end{eqnarray*} \\
\hline
\begin{eqnarray*}
\nonumber \\[-6ex]
\textbf{\textrm{min}} \;\quad \sum_{e \in E} c_{e} y_{e} \quad\quad \label{eq:LPgoalR}\\\\[-2ex]
    \nonumber\\
\textbf{\textrm{s.t.}} \quad d_{r} \sum_{e \in s} y_{e} \geq v_{r} &\forall r \in \mathcal{R},\forall s \in S_{r}\label{eq:LPvaluesR}\\
    y_{e} \geq 0 \quad\quad\quad& \forall e \in E \label{eq:LPedgesR}
\\[-7ex] \nonumber
\end{eqnarray*}\\
\hline
\end{tabular}
\caption{\label{cap:UFPRepeatILPandDualLp} The integer linear
program of the unsplittable flow with repetitions problem (top),
and the dual of its relaxation (bottom). Note that $S_{r}$ denotes
the set of all the simple paths between $s_{r}$ and $t_{r}$ in
$G$, $S = \bigcup_{r \in \mathcal{R}} S_{r}$, and $d_{s}$ and
$v_{s}$ denote the respective demand and value of path $s$.}
\end{figure}

We now turn to analyze algorithm \textsf{Bounded-UFP-Repeat}. For
the sake of simplicity, we shall use the same notation, which was
introduced in Subsection~\ref{cha:UfpAlgoSect}, with one
exception. Namely, since there are no $z$-type variables in the
dual linear program, we let $D(i) = \sum_{e \in E} c_{e}
y_{e}^{i}$ denote the value of the dual solution at the end of the
$i$-th iteration, and neglect $D_{1}(i)$ and $D_{2}(i)$. Notice
that Lemma~\ref{th:UfpFeasibility} is applicable also in this case
and thus, the correctness of the algorithm follows. Consequently,
in the sequel, we prove that the algorithm under consideration
achieves an approximation ratio of $(1 + \epsilon)$. We begin by
establishing a analogous claim to Claim~\ref{th:AlphaBound}, which
upper bounds $\alpha(i)$.

\begin{claim} \label{th:AlphaBoundRepeat}
$\alpha(i) \leq \frac{D(i)}{D}$, in every iteration $i \geq 0$.
\end{claim}
\begin{proof}
Consider the $(i + 1)$-th iteration, and let $p$ denote the path,
which is selected in this iteration. The path $p$ corresponds to a
request such that $\frac{d_{p}}{v_{p}} \sum_{e \in p}y_{e}^{i}$ is
minimal. Namely, every other path $p'$, which corresponds to a
request, satisfies
$$
\frac{d_{p'}}{v_{p'}} \sum_{e \in p'} y_{e}^{i} \geq
\frac{d_{p}}{v_{p}} \sum_{e \in p}y_{e}^{i} = \alpha(i) \text{, thus
} d_{p'} \sum_{e \in p'} \frac{y_{e}^{i}}{\alpha(i)}
\geq v_{p'} \ .
$$
This implies that if we multiply $y_{e}^{i}$ by $\alpha(i)^{-1}$,
for every $e \in E$, then all the dual linear program constraints
become satisfied, i.e. the modified variables constitute a
feasible fractional solution to the dual linear program. Hence, $D
\leq D(i) \alpha(i)^{-1}$.~
\end{proof}

Next, we prove that the algorithm achieves the claimed $(1 +
\epsilon)$-approximation for the problem under consideration.

\begin{lemma} \label{th:UfpApproxRepeat}
\textsf{Bounded-UFP-Repeat}$(\epsilon)$ returns an $(1 +
6\epsilon)$-approximate solution for the $\frac{\ln
m}{\epsilon^{2}}$-bounded unsplittable flow with repetitions
problem, for any $\epsilon \in (0,\frac{1}{6}]$.
\end{lemma}
\begin{proof}
For every $i \geq 0$, one can derive that
\setlength\arraycolsep{2pt}
\begin{eqnarray*}
D(i + 1) & \leq & D(i) + B \epsilon (1 + \epsilon) \cdot \Delta P(i + 1) \cdot \alpha(i) \\
& \leq & D(i) + B \epsilon (1 + \epsilon) \cdot \Delta P(i+1) \cdot \frac{D(i)}{D} \\
& = & D(i) \Big(1 + B \epsilon (1 + \epsilon) \frac{\Delta  P(i+1)}{D}\Big) \\
& \leq & D(i) e^{\big(B \epsilon (1 + \epsilon) \frac{\Delta
P(i+1)}{D}\big)} \ .
\end{eqnarray*}
The first inequality follows from Claim~\ref{th:D1DeltaBound},
while noticing that $D(i)$ may replace $D_{1}(i)$. The second
inequality is due to Claim~\ref{th:AlphaBoundRepeat}. Finally, the
last inequality results from the fact that $1 + a \leq e^{a}$.
This implies that
\begin{eqnarray*}
D_{1}(i + 1) & \leq & D_{1}(0) e^{\big(B \epsilon (1 + \epsilon) \sum_{j=0}^{i} \frac{\Delta P(j + 1)}{D}\big)} \\
& \leq & e^{\big(B  \epsilon^{2} + B \epsilon (1 + \epsilon) \sum_{j=0}^{i} \frac{\Delta P(j + 1)}{D}\big)} \ ,
\end{eqnarray*}
where the first inequality results from the expansion of the
recursion, and the second inequality follows from $D_{1}(0) \leq
e^{B \epsilon^{2}}$, which dues to $D_{1}(0) = m$, and $B \geq
\frac{\ln m}{\epsilon^{2}}$. Lets assume that the algorithm
terminates after $t + 1$ iterations. Accordingly, inspecting the
stoping condition of the main loop, we attain that
$$
e^{\epsilon (B-1)} < D_{1}(t + 1) \leq e^{\big(B  \epsilon^{2} + B
\epsilon (1 + \epsilon) \sum_{j=0}^{t} \frac{\Delta P(j +
1)}{D}\big)} \ .
$$
This can be simplified to $\frac{\epsilon (B-1) - B
\epsilon^{2}}{B \epsilon (1 + \epsilon)} < \sum_{j=0}^{t}
\frac{\Delta P(j + 1)}{D}$. Notice that $\sum_{j=0}^{t} \Delta P(j
+ 1)$ is a telescopic sum that is equal to $P(t + 1) - P(0) = P$,
where the equality dues to $P(t + 1) = P$ and $P(0) = 0$. In
addition, notice that $1 - 3 \epsilon \leq \frac{\epsilon (B-1) -
B \epsilon^{2}}{B \epsilon (1 + \epsilon)}$ for any positive
$\epsilon$. Consequently, we obtain $1 - 3\epsilon \leq
\frac{P}{D}$. Simplifying this expression even further yields
$\frac{D}{P} \leq \frac{1}{1 - 3\epsilon} \leq 1 + 6\epsilon$,
which by the weak LP duality establishes the lemma.~
\end{proof}

We are now ready to prove the main theorem of this subsection.
\begin{exrefproof}{Theorem}{\ref{th:UfpRepeatResult}}
As noted before, the correctness of the algorithm directly follows
from Lemma~\ref{th:UfpFeasibility}. In addition, the approximation
guarantee is proved in Lemma~\ref{th:UfpApproxRepeat}. We now turn
to argue that the running time of the algorithm is polynomial in
$m$ and $\frac{c_{\max}}{d_{\min}}$, where $c_{\max} = \max_{e} \{
c_{e} \}$, and $d_{\min} = \min_{r} \{ d_{r} \}$. Consider some
edge $e \in E$. Recall that $y_{e}^{0} = \frac{1}{c_{e}}$ and
$y_{e}^{t} \leq \frac{1}{c_{e}}e^{\epsilon B}$, where $t$ denotes
the index of the last iteration of the algorithm. In addition,
notice that every time that the algorithm increments $y_{e}$, it
is by a multiplicative factor of at least $e^{\epsilon B d_{\min}
/ c_{\max}}$. Consequently, the number of iterations in which
$y_{e}$ is incremented is at most $\frac{c_{\max}}{d_{\min}}$.
This implies that the running time of the algorithm is bounded by
$m \frac{c_{\max}}{d_{\min}}$, as there are $m$ edges.~
\end{exrefproof}

% REFERENCES %%%%%%%%%%%%%%%%%%%%%%%%%%%%%%%%%%%%%%%%%%%%%%%%%%%%%%%
%\bibliographystyle{abbrv}
%\bibliography{TruthfulUFP}

\vfill\eject

% APPENDIX %%%%%%%%%%%%%%%%%%%%%%%%%%%%%%%%%%%%%%%%%%%%%%%%%%%%%%%%%
%\pagebreak
%\begin{appendix}
%\section{Unsplittable Flow Problem} \label{cha:UfpApxSect}
%In this section, we establish all the omitted proofs of
%Section~\ref{cha:UfpSect}.

%\subsection{The algorithm analysis} \label{cha:UfpAlgoApxSect}

%In what follows, we prove that the algorithm outputs a feasible
%solution. Namely, we demonstrate that the edges capacity
%constraints are not violated.

%Next, we establish the monotonicity and exactness of the
%algorithm.

%We now turn to prove the three claims, which were used to
%demonstrate the approximation guarantee of the algorithm.

%We are now ready to prove the main result of this subsection.

%\subsection{Inapproximability result} \label{cha:UfpApxInapprox}

%\section{Single-minded Multi-unit \\ Combinatorial Auction} \label{cha:MucaApxSect}

%\subsection{Inapproximability result}

%\section{Unsplittable Flow with \\ Repetitions Problem} \label{cha:UfpRepeatApxSect}

%\subsection{The algorithm analysis}
%In this section, we prove Theorem~\ref{th:UfpRepeatResult}, which
%is the main finding of Section~\ref{cha:UfpRepeatSect}.

%\end{appendix}

\end{document}